\documentclass[a4paper,fleqn]{cas-sc}
\usepackage{soul}
\usepackage[sort&compress,numbers]{natbib}
\usepackage{lineno}
\usepackage{setspace}
\usepackage{verbatim}
\usepackage{graphics}
\usepackage{graphicx}
\usepackage{cuted}
\usepackage{lipsum}
\usepackage{multirow}

\def\tsc#1{\csdef{#1}{\textsc{\lowercase{#1}}\xspace}}
\tsc{WGM}
\tsc{QE}
\tsc{EP}
\tsc{PMS}
\tsc{BEC}
\tsc{DE}

\begin{document}

\let\WriteBookmarks\relax
\def\floatpagepagefraction{1}
\def\textpagefraction{.001}
\shorttitle{CO$_2$ Adsorption on Biphenylene Network}
\shortauthors{Tromer \textit{et~al}.}

\title [mode = title]{Unveiling the CO$_2$ Adsorption Capabilities of Biphenylene Network Monolayers through DFT Calculations}

\author[1]{K. A. Lopes Lima}
\author[1,2]{L. A. Ribeiro Junior}
\cormark[1]
\ead{ribeirojr@unb.br}

\address[1]{University of Bras\'{i}lia, Institute of Physics, 70.910-900, Bras\'{i}lia, Brazil.}
\address[2]{Computational Materials Laboratory, LCCMat, Institute of Physics, University of Bras\'ilia, 70910-900, Bras\'ilia, Brazil}


\begin{abstract}
Nanomaterial synthesis and characterization advancements have led to the discovery of new carbon allotropes, such as the biphenylene network (BPN). BPN consists of four-, six-, and eight-membered rings of sp$^2$-hybridized carbon atoms. Here, we employ density functional theory (DFT) calculations to investigate the CO$_2$ adsorption capabilities in pristine and vacancy-endowed BPN monolayers. Our findings indicate that BPN lattices with a single-atom vacancy exhibit higher CO$_2$ adsorption energies than pristine BPN. Unlike other 2D carbon allotropes, BPN does not exhibit precise CO$_2$ sensing and selectivity by altering its band structure configuration. In pristine lattices, CO$_2$ molecules are physisorbed in the eight-membered rings, while defective regions of vacancy-endowed lattices enable chemisorption of CO$_2$. Regarding CO$_2$ physisorption, the recovery time values are minimal, suggesting a rapid interaction between BPN and this molecule, with negligible relaxation time required to alter the electronic properties of BPN lattices.
\end{abstract}



\begin{keywords}
Biphenylene Network \sep Oxygen Molecule \sep Vacancy \sep Adsorption \sep Density Functional Theory.
\end{keywords}

\maketitle
\doublespacing

\section{Introduction}

Two-dimensional (2D) carbon allotropes have garnered considerable attention due to their unique properties and wide-ranging applications across various fields \cite{huang2021two,tang2014two}. Comprised solely of carbon atoms, they typically possess an ultra-thin thickness of just one atom, leading to distinct properties that set them apart from their 1D and 3D counterparts \cite{chen2010graphene}. 2D carbon allotropes exhibit exceptional structural characteristics, offering versatility and customization for applications in flat electronics \cite{khan2020recent,jia2017synthesis}. Moreover, their atomic-level structural modifications enable precise tailoring of their properties to meet specific application requirements \cite{bhimanapati2015recent}.

Graphene, a single layer of carbon atoms arranged in a hexagonal lattice, is one of the most renowned 2D carbon allotropes \cite{geim2007rise}. Its excellent electrical conductivity \cite{wu2012graphene}, impressive mechanical strength \cite{papageorgiou2017mechanical}, and thermal stability \cite{nan2013thermal} position it as a promising material for a diverse range of applications in electronics \cite{avouris2010graphene}, energy storage \cite{raccichini2015role}, and other fields \cite{shen2012biomedical,wan2011graphene}. However, the absence of a band gap in graphene poses a hurdle for specific electronic applications like transistors or digital logic circuits, which necessitate materials with semiconducting band gaps \cite{schwierz2010graphene,weiss2012graphene}. Consequently, researchers have explored various approaches to introduce a band gap into graphene \cite{zhou2007substrate} or develop novel materials that retain its unique properties while possessing a well-defined band gap \cite{desyatkin2022scalable,hou2022synthesis,meirzadeh2023few}.

Recently, various novel 2D carbon allotropes have emerged, significantly expanding the pool of available materials for potential applications \cite{desyatkin2022scalable,hou2022synthesis,meirzadeh2023few,fan2021biphenylene,pan2023long,toh2020synthesis,hu2022synthesis}. These include monolayer amorphous carbon \cite{toh2020synthesis}, monolayer fullerene network \cite{hou2022synthesis,meirzadeh2023few}, gamma-graphyne \cite{desyatkin2022scalable,hu2022synthesis}, and the biphenylene network (BPN) \cite{hou2022synthesis}. Of particular interest is the BPN, an sp$^2$-hybridized carbon allotrope boasting a flat structure with a well-organized arrangement of four-, six-, and eight-membered rings of carbon atoms \cite{hou2022synthesis}. Its synthesis involved an on-surface interpolymer dehydrofluorination reaction, resulting in an exceptionally flat structure. Notably, the BPN exhibits metallic behavior, rendering it suitable for applications such as a conducting wire or anode material in lithium-ion batteries, offering promising prospects for its usage \cite{liu2022electronic,chen2023new,ferguson2017biphenylene,mu2023superconducting}.

BPN offers a fresh perspective for exploring non-graphene sp$^2$ carbon allotropes and their distinctive characteristics. Notably, BPN possesses eight-membered rings that form pores capable of efficiently trapping and adsorbing small molecules like CO$_2$ and methane \cite{esfandiarpour2022carbon,su2022theoretical}. This attribute positions BPN as a promising contender for devising strategies to mitigate the adverse impacts of increasing atmospheric CO$_2$ levels. Although many nanostructured materials have been explored for their CO$_2$ adsorption capabilities \cite{mishra2011carbon,cabrera2009adsorption,takeuchi2017adsorption}, studies in the literature have yet to investigate the potential of BPN for CO$_2$ adsorption.

In this study, we employed density functional theory (DFT) calculations to investigate the adsorption capability of both pristine and vacancy-endowed BPN monolayers regarding CO$_2$. BPN lattices with a single-atom vacancy exhibit higher CO$_2$ adsorption energies than pristine counterparts. Unlike graphene and other 2D carbon allotropes, BPN lattices do not rely on changes in their band structure configurations for precise CO$_2$ sensing and selectivity. 

\section{Methodology}

Our study aimed to investigate the impact of CO$_2$ adsorption on the electronic and structural properties of pristine and vacancy-endowed BPN lattices through DFT calculations. We used the DMol3 code \cite{delley_JCP,delley_JCP2} within the Biovia Materials Studio software \cite{systemes2017biovia} to perform these calculations. For a proper treatment of van der Waals interactions, we employed the Grimme scheme \cite{grimme2006} and employed the Perdew-Burle-Ernzerhof (PBE) functional \cite{kresse1999,perdew1996,grimme2006} with unrestricted spin (DNP) within the framework of the exchange-correlation generalized gradient approximation (GGA). We incorporated semi-core DFT pseudopotentials to accurately describe the interactions between valence electrons and atomic nuclei and employed a numerical basis set of atomic orbitals with polarized functions. Additionally, we accounted for the basis set superposition error (BSSE) correction \cite{delley2002} using the counterpoise method to ensure the precision and reliability of our results.

\begin{figure*}[pos=ht!]
	\centering
	\includegraphics[width=\linewidth]{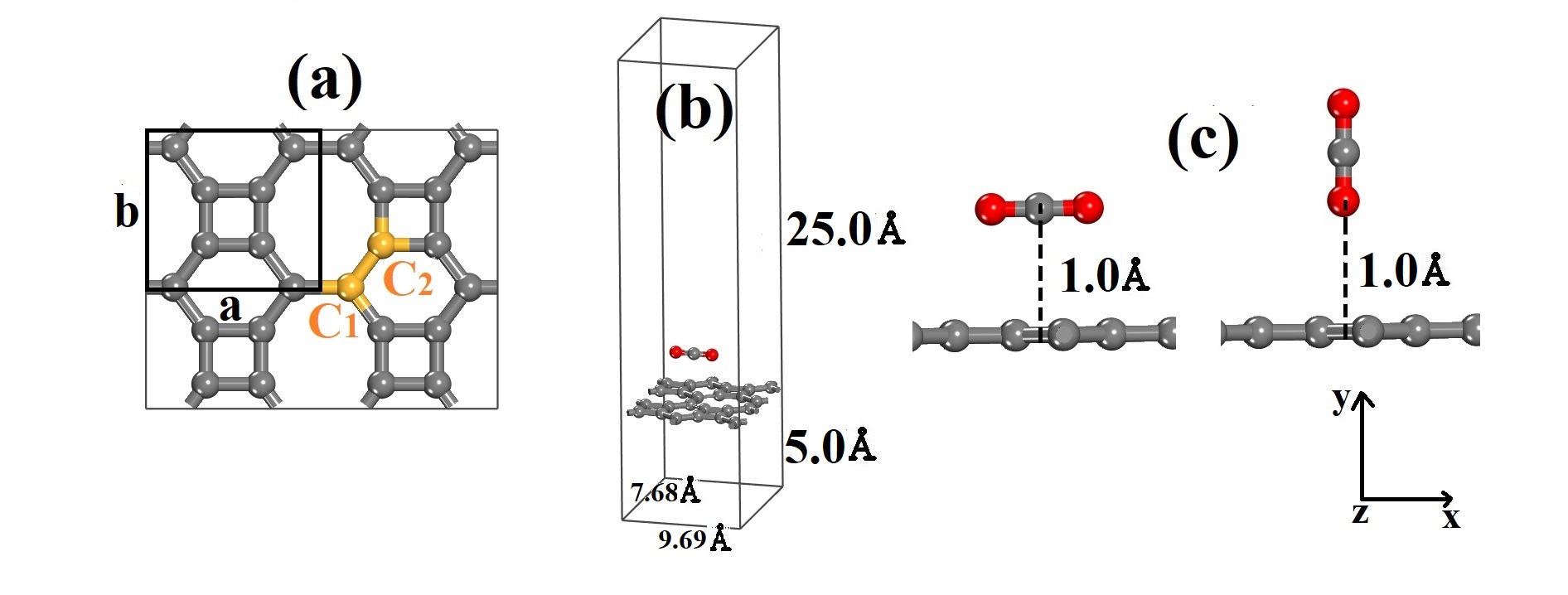}
	\caption{In panel (a), a schematic representation of the optimized pristine BPN sheet is presented, which includes the unit cell with lattice vectors $a$ and $b$. In yellow, the atoms C$_1$ and C$_2$ represent the lattice sites where vacancies are imposed. The initial configuration for the BPN/CO$_2$ system is presented in panel (b). Panel (c) shows the side view of the BPN/CO$_2$ system, highlighting the initial distance between CO$_2$ (in its horizontal and vertical alignments) and the BPN sheet. Oxygen and carbon atoms are red and grey, respectively.} 
	\label{fig1}
\end{figure*}

To optimize the system's geometry, we implemented several parameters to ensure the accuracy of our calculations. A $5\times 5\times 2$ Monkhorst-Park k-point mesh in the Brillouin zone \cite{hendrik1976} was used. Additionally, we set a self-consistent field (SCF) tolerance of $1\times10^{-5}$ eV/atom and established a maximum force on each atom of $0.001$ Ha/\AA. A maximum value of $0.005$ \r{A} was adopted for the atomic displacement. We employed a vacuum spacing of 30 \r{A} in lattices with dimensions of $2\times2\times1$ to prevent interactions between layer images. Moreover, cutoff energy of 450 eV and a $10\times 10\times 2$ Monkhorst-Park k-point were employed. It is important to note that these parameters have previously proven successful in studying small molecule adsorption on nanostructured surfaces \cite{lima2021,tlima2019,paura2014,mananghaya2020adsorption,lima2019}.

Figure \ref{fig1}(a) illustrates the unit cell of the optimized BPN monolayer featuring lattice vectors $a$ and $b$. The BPN structure consists of a unit cell with $6$ atoms and belongs to the $P1$ plane group, with lattice parameters measuring $4.26$ \r{A} and $3.88$ \r{A}, which aligns with previous findings \cite{bafekry2021biphenylene,mortazavi2022anisotropic,rahaman2017metamorphosis}. The initial configuration of the BPN/CO$_2$ systems is presented in Figure \ref{fig1}(b). For clarity, Figure \ref{fig1}(c) provides a side view of the initial configurations, highlighting the initial distance of 1.0 \r{A} between CO$_2$ and the BPN sheet, with CO$_2$ positioned in both horizontal and vertical alignments.

\section{Results}

We initiate our discussion by presenting the optimized configurations for all the BPN@CO$_2$ systems examined in this study, following the representation in Figures \ref{fig1}(b-c). The outcomes for the pristine case (BPN@CO$_2$), C$_1$ vacancy case (BPN-DEF/C$_1$@CO$_2$), and C$_2$ vacancy case (BPN-DEF/C$_2$@CO$_2$) with the initial orientations of the CO$_2$ molecule in both the vertical (V) and horizontal (H) directions are exhibited in Figures \ref{fig02}(a)-(f). DEF denotes defective lattices.

\begin{figure*}[pos=ht!]
	\centering
	\includegraphics[width=\linewidth]{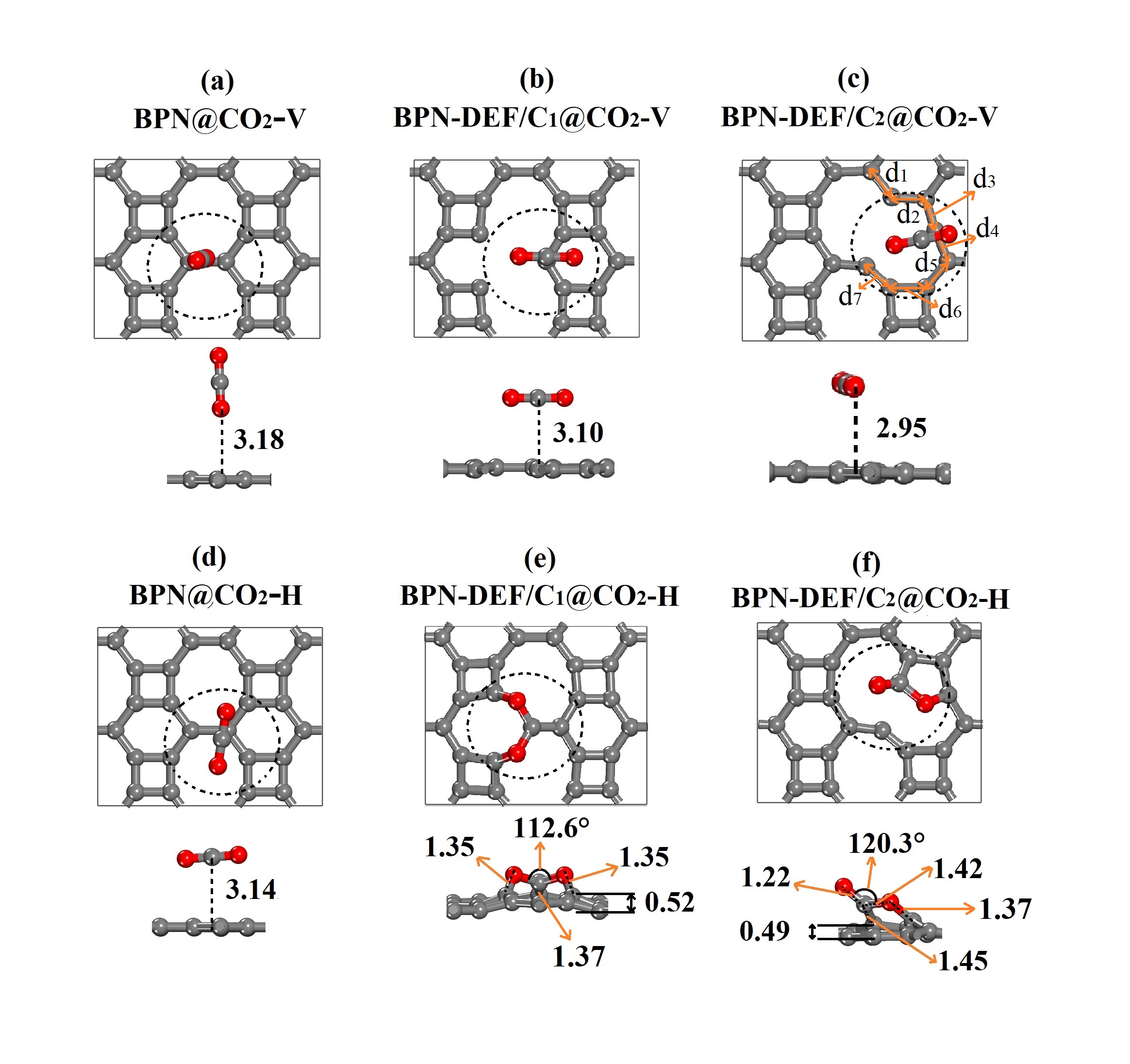}
	\caption{(a-f) show schematic representations of the optimized configurations for the BPN/CO$_2$ cases: (a-f) BPN@CO$_2$-V, BPN-DEF/C$_{1}$@CO$_2$-V, BPN-DEF/C$_{2}$@CO$_2$-V, BPN@CO$_2$-H, BPN-DEF/C$_{1}$@CO$_2$-H, BPN-DEF/C$_{2}$@CO$_2$-H, respectively. Oxygen and carbon atoms are red and grey, respectively. In panel (c), the orange arrows stand for the novel C-C bonds (d$_1$, d$_2$, d$_3$, d$_4$, d$_5$, d$_6$, and d$_7$) formed after bond reconfiguration imposed by the vacancy type C$_2$. DEF denotes defective lattices.}
	\label{fig2}
\end{figure*}

The initial position of the CO$_2$ molecule underwent displacement in all the examined cases, resulting in adsorption distances of $3.18$ \r{A}, $3.10$ \r{A}, $2.95$ \r{A}, $3.14$ \r{A}, $0.52$ \r{A}, and $0.49$ \r{A} for BPN@CO$_2$-V, BPN-DEF/C$_1$@CO$_2$-V, BPN-DEF/C$_2$@CO$_2$-V, BPN@CO$_2$-H, BPN-DEF/C$_1$@CO$_2$-H, and BPN-DEF/C$_2$@CO$_2$-H, respectively. Among all the cases examined, only the pristine case (BPN@CO$_2$-V, Figure \ref{fig2}(a)) demonstrated a minimal change in the initial orientation of the vertically positioned molecule following structure optimization. In the cases shown in Figures \ref{fig2}(b-d), the CO$_2$ molecule maintained its horizontal alignment relative to the BPN sheet throughout the optimization process. These interactions between BPN and CO$_2$ in Figures \ref{fig2}(a-d) are characterized by a physisorption mechanism during adsorption.  

Conversely, Figures \ref{fig2}(e) and \ref{fig2}(f) demonstrate that BPN and CO$_2$ interact through a chemisorption mechanism. This adsorption type disrupts the BPN lattice locally, forming a localized buckling with a magnitude of approximately 0.5 \r{A}. In Figure \ref{fig2}(e), all CO$_2$ atoms establish bonds with BPN atoms within a range of 1.35-1.37 \r{A}, inducing a new angle in the CO$_2$ molecule of 112.6$^\circ$. In Figure \ref{fig2}(f), only two atoms from the CO$_2$ molecule bind to the BPN, forming bonds with lengths ranging from 1.37-1.45 \r{A}. In this case, the molecule's resulting angle is 120.3$^\circ$.

In the BPN@CO$_2$-V and BPN@CO$_2$-H cases (Figures \ref{fig2}(a) and \ref{fig2}(d), respectively), the CO$_2$ molecule was adsorbed in proximity to two adjacent octagonal rings, indicating an adsorption mechanism in pristine BPN that is independent of molecule orientation. On the other hand, in the defective cases BPN-DEF/C$_1$@CO$_2$-V and BPN-DEF/C$_2$@CO$_2$-V (Figures \ref{fig2}(b) and \ref{fig2}(c), respectively), CO$_2$ was adsorbed peripherally to the defect, resulting in similar adsorption distances. 

In contrast to previous DFT studies on CO$_2$ adsorption on graphene \cite{sanyal2009molecular}, defects in BPN have minimal impact on the adsorption distance or CO$_2$ orientation. This feature can be attributed to the presence of the eight-atom ring in the BPN structure, which acts as a pore and exhibits a slightly weaker interaction potential than a single-atom vacancy in the lattice. In the case of graphene, for example, a twelve-atom ring resulting from a single-atom vacancy forms a larger pore diameter than the other hexagonal rings in its structure \cite{sanyal2009molecular}. For the C$_2$ vacancy cases, the pore region undergoes bond reconfiguration (see Figures \ref{fig2}(c) and \ref{fig2}(f)). The new C-C bonds, indicated by the orange arrows in Figure \ref{fig2}(c), are d$_1=$1.42, d$_2=$1.45, d$_3=$1.40, d$_4=$1.39, d$_5=$1.48, d$_6=$1.52, and d$_7=$1.37.

\begin{figure*}[pos=ht!]
	\centering
	\includegraphics[width=\linewidth]{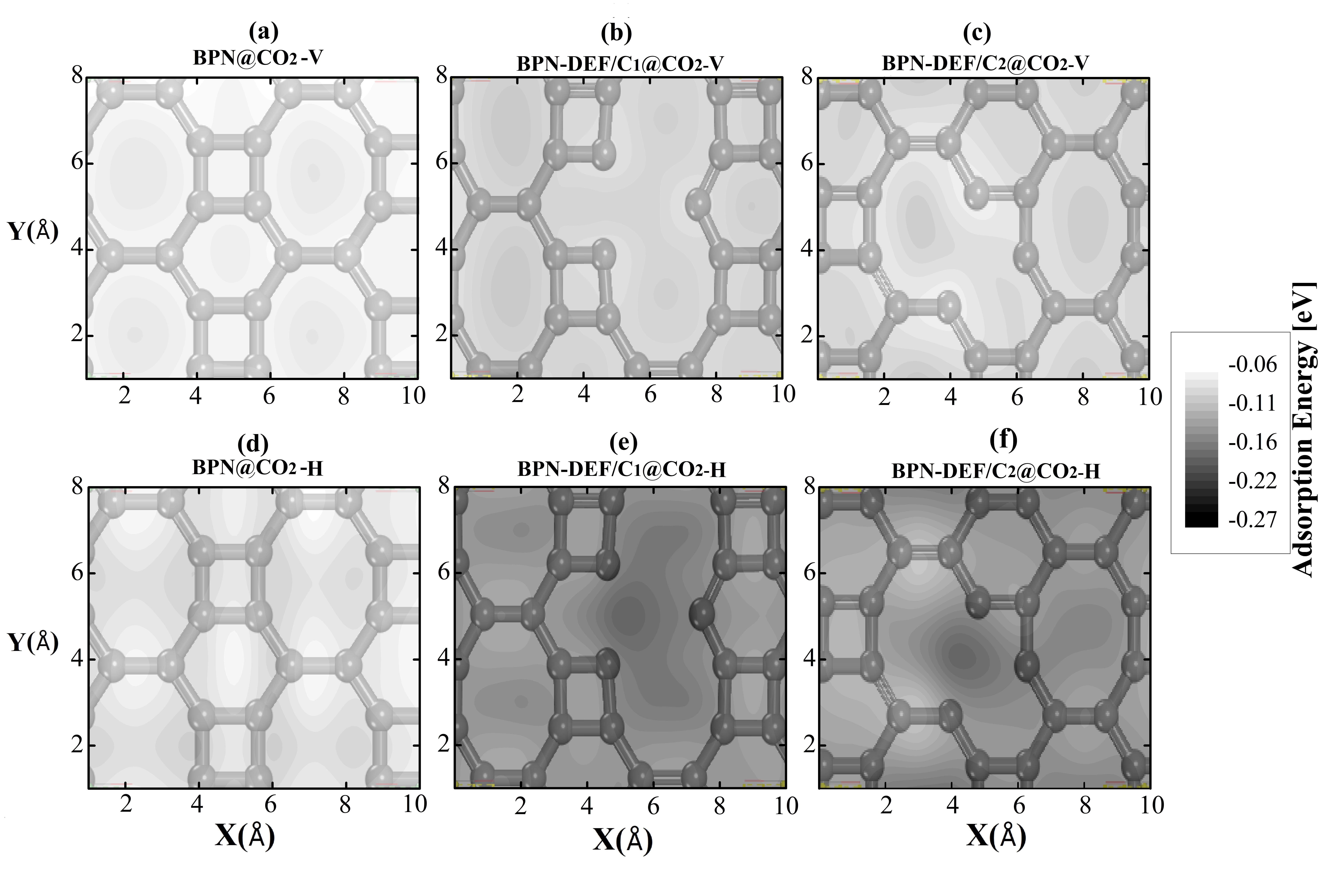}
	\caption{Adsorption energy maps for the minimum energy configurations in the (a) BPN@CO$_2$-V, BPN-DEF/C$_{1}$@CO$_2$-V, (b) BPN-DEF/C$_{2}$@CO$_2$-V, (c) BPN@CO$_2$-H, (d) BPN-DEF/C$_{1}$@CO$_2$-H, and (f) BPN-DEF/C$_{2}$@CO$_2$-H cases.}
    \label{fig3}
\end{figure*}

To comprehensively analyze the adsorption energies of the BPN/CO$_2$ complexes, we have generated adsorption energy maps for all cases, as shown in Figure \ref{fig3}. These maps were created by systematically varying the position of the CO$_2$ molecule along the x and y directions of the BPN surface, with a step size of 0.5 \r{A}. The CO$_2$ molecule was positioned 3.0 \r{A} above the BPN surface in these calculations. A clear trend emerges, with the regions inside the octagonal rings displaying lower adsorption energies, ranging from -0.22 eV to -0.08 eV. Furthermore, Figure \ref{fig3} emphasizes that interaction energies are significantly lower in regions with lattice defects and where the CO$_2$ molecule is aligned horizontally, ranging from -0.22 eV to -0.27 eV.

In the BPN@CO$_2$-V case (Figure \ref{fig3}(a)), the center of the octagonal rings exhibits the lowest adsorption energies, indicating stronger interaction between BPN and an oxygen atom in the CO$_2$ molecule. In the BPN@CO$_2$-H case (Figure \ref{fig3}(d)), the regions with the lowest adsorption energies extend towards the C-C bonds due to the proximity of the carbon atom in the molecule to the BPN plane. In these cases, the adsorption energies for CO$_2$ range from -0.11 eV to -0.06 eV.

The BPN lattice with a C$_1$ vacancy type shows the largest pore region among all the structures studied, as depicted in Figures \ref{fig3}(b) and \ref{fig3}(e). The CO$_2$ adsorption energies for the BPN-DEF/C$_1$@CO$_2$-V and BPN-DEF/C$_1$@CO$_2$-H cases range from -0.22 eV to -0.08 eV. In the DEF/C$_2$@CO$_2$-V case, the adsorption energies vary from -0.16 eV to -0.08 eV. The lowest CO$_2$ adsorption energy (-0.25 eV) is observed in the BPN-DEF/C$_2$@CO$_2$-H case, as shown in Figure \ref{fig3}(f). 

\begin{figure*}[pos=ht!]
	\centering
	\includegraphics[width=\linewidth]{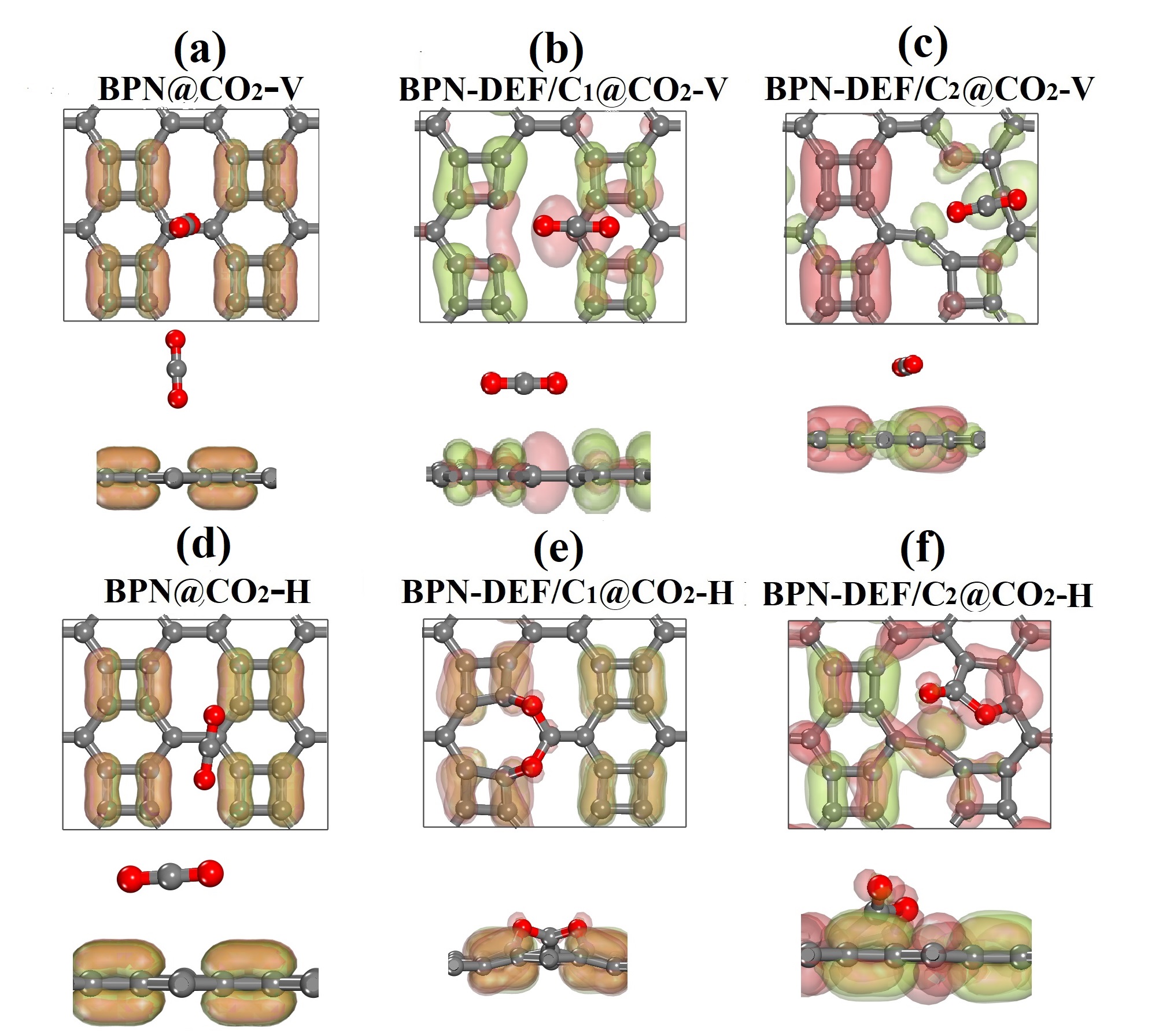}
	\caption{Schematic representation for the distribution of the Highest Occupied Molecular Orbital (HOMO, in green) and Lowest Unoccupied Molecular Orbital (LUMO, in red) configurations.}
    \label{fig4}
\end{figure*}

To visualize the localization of the Highest Occupied Molecular Orbital (HOMO, in green) and Lowest Unoccupied Molecular Orbital (LUMO, in red), we present Figure \ref{fig4}. Importantly, the HOMO and LUMO orbitals provide valuable insights into the electronic structure of the nanostructure and the adsorbed molecule. By visualizing these orbitals, we can determine the distribution and localization of electronic states, identify bonding and antibonding regions, and assess the interaction between the molecule and the nanostructure. Generally, the HOMO and LUMO orbitals span across the BPN lattice. In the pristine case, they are localized on the C-C bonds forming the square ring, as depicted in Figures \ref{fig4}(a) and \ref{fig4}(d).

With the presence of a single-atom vacancy, the well-ordered distribution pattern of the HOMO and LUMO orbitals observed in the pristine case is disrupted, and these orbitals are scattered randomly around the vacancy region, as shown in Figures \ref{fig4}(b,c,f). Interestingly, for the case BPN-DEF/C$_1$@CO$_2$-H (see Figure \ref{4}(e)), the binding of the CO$_2$ molecule to the BPN lattice leads to reforming the previously defective octagon. In this process, the distribution of the HOMO and LUMO orbitals undergoes a transition, resembling the configuration observed in the pristine case. In chemisorption cases (see Figures \ref{fig4}(e) and \ref{fig4}(f)), a portion of the LUMO orbital is found on the CO$_2$ molecule. This process results in a slight charge transfer from BPN to CO$_2$, as indicated in Table \ref{tab1}.

Next, we focus on the recovery time ($\tau$), representing the duration required for stabilizing adsorption. A higher $\tau$ value indicates better adsorption performance. We calculate $\tau$ using the equation $\tau=\nu^{-1}\times \exp(-E_\textrm{ads}/k_BT)$, where $\nu$ corresponds to the oscillation frequency of the molecule ($10^{12}$ $s^{-1}$ \cite{timsorn_IOP}), $k_B$ is the Boltzmann constant, and $T$ denotes the temperature (298 K). 

\begin{table*}[pos=ht!]
 \centering
 \caption{Calculated adsorption energies (E$_{ads}$), amount of transferred charge from CO$_2$ to BPN (Q$_t$), and the recovery time $\tau (s)$.}
\begin{tabular}{*{7}{l}}
\hline
Structures         & E$_\textrm{ads}$ (eV)   & Q$_{t}$ ($e$)  & $\tau$ (s) \\
\hline
BPN                                 & $---$        & $---$       & $---$  \\
BPN@CO$_{2}$-V                      & $-0.22$      &$-0.001$     & $5.25\times10^{-9}$   \\
BPN@CO$_{2}$-H                      & $-0.31$      &$-0.001$     & $1.74\times10^{-7}$   \\
BPN-DEF/C$_{2}$@CO$_{2}$-V          & $-0.54$      &$-0.003$     & $1.35\times10^{-3}$   \\
BPN-DEF/C$_{1}$@CO$_{2}$-V          & $-0.65$      &$-0.005$     & $9.80\times10^{-2}$   \\
BPN-DEF/C$_{1}$@CO$_{2}$-H          & $-3.52$      &$-0.47$      & $---$   \\
BPN-DEF/C$_{2}$@CO$_{2}$-H          & $-3.60$      &$-0.25$      & $---$   \\
\hline
\label{tab1}
\end{tabular}
\end{table*}

Upon examining Table \ref{tab1}, we observe that both $\tau$ and the transferred charge ($Q_t$) increase with the adsorption energy ($E_\textrm{ads}$). Table \ref{tab1} summarizes the $E{ads}$, $Q_t$, and $\tau$ values obtained for each case. The system with the longest recovery time is BPN-DEF/C$_1$@CO$_2$-V, approximately $9.80\times 10^{-2}$ s. Conversely, the pristine case BPN@CO$_2$-V exhibits the shortest recovery time, around $5.25\times 10^{-9}$ s. These relatively small $\tau$ values suggest that the interaction between BPN and CO$2$ occurs rapidly, without a significant relaxation time required to alter the BPN electronic band structure, as discussed below. 

\begin{figure*}[pos=ht!]
	\centering
	\includegraphics[width=\linewidth]{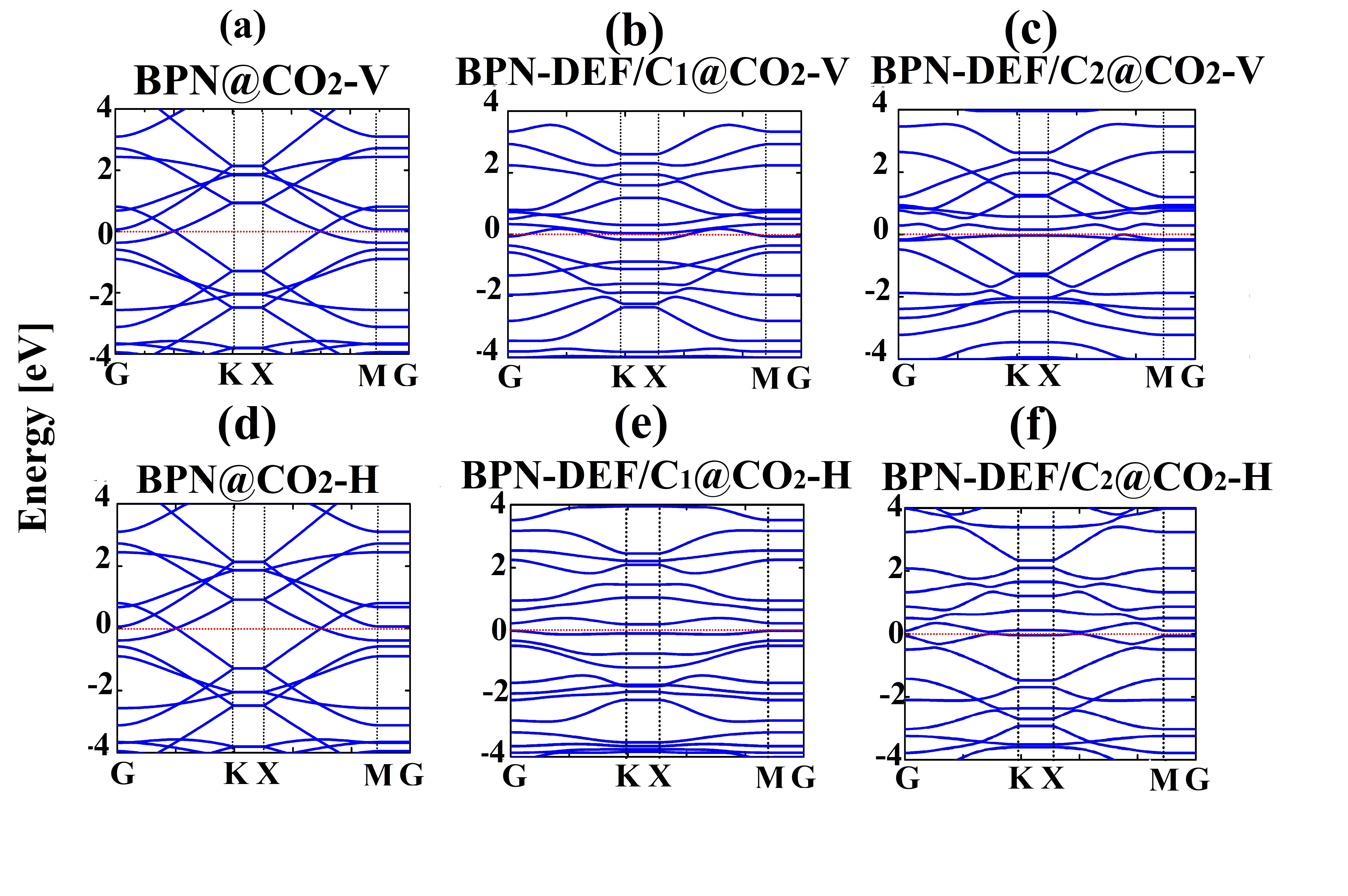}
	\caption{The electronic band structures for all the cases studied here. The dashed red line represents the Fermi level.}
    \label{fig5}
\end{figure*}

Finally, Figure \ref{fig5} showcases the electronic band structure for all the cases discussed here. The dashed red line denotes the Fermi level. Previous studies on the electronic band structure of pristine BPN can be found in references \cite{bafekry2021biphenylene,mortazavi2022anisotropic,rahaman2017metamorphosis}. Pristine BPN displays a metallic signature with a Dirac cone in the middle of the band along the G-K and X-M paths. The band structure configurations for isolated pristine BPN and vacancy-endows BPN lattices are presented in the Supplementary Material.

The presence of a single-atom vacancy disrupts this band profile, resulting in the elimination of the Dirac cones and the opening of a small bandgap of approximately 100 meV. Interestingly, BPN does not exhibit precise CO$_2$ sensing and selectivity by altering its band structure configuration. The adsorption of CO$_2$ does not significantly alter the band dispersion profiles of the BPN lattices, as evidenced by their low energy adsorption and quick recovery times. Unlike other 2D carbon allotropes \cite{sanyal2009molecular,takeuchi2017adsorption,cabrera2009adsorption,mmishra2011carbon}, BPN lattices do not exhibit precise CO$_2$ sensing and selectivity based on changes in their band structure configurations.

\section{Conclusions}

Our study employed DFT calculations to analyze the CO$_2$ adsorption characteristics of pristine and vacancy-endowed BPN monolayers. The computational approach involved generating potential energy maps to identify specific regions of the BPN surface with lower CO$_2$ adsorption energies.

The obtained adsorption distances and energy values collectively suggest CO$_2$ adsorption through physisorption and chemisorption mechanisms. In the pristine BPN cases, we observed that the CO$_2$ molecule was adsorbed near two octagonal rings, implying an adsorption mechanism independent of the molecule's orientation. Conversely, in the defective cases with vertically aligned CO$_2$, this molecule was adsorbed peripherally to the defect, resulting in similar adsorption distances. 

Defects in BPN have a negligible impact on the adsorption distance, but they are sensitive to the CO$_2$ orientation. This distinction can be attributed to an eight-atom ring within the BPN structure, which functions as a pore with a slightly weaker interaction potential than a single-atom vacancy in the lattice. The physisorption process prevailed when the CO$_2$ molecule was aligned vertically concerning the BPN surface. However, when the CO$_2$ molecule was aligned horizontally on the BPN surface, the adsorption process was dominated by chemisorption. 

A clear trend emerged in analyzing the adsorption maps. The adsorption energies tend to present suitable values in regions inside the octagonal rings, ranging from -0.22 eV to -0.08 eV. However, significantly lower interaction energies were observed within the lattice defects for horizontally aligned CO$_2$ molecules, about -0.25 eV.

The band structure of pristine BPN exhibits a metallic signature with Dirac cones in the middle of the band. However, a single-atom vacancy disrupts this band profile, eliminating the Dirac cones and resulting in a small bandgap opening of approximately 100 meV. Importantly, BPN does not exhibit precise CO$_2$ sensing and selectivity by altering its band structure configuration. 

\section*{Acknowledgements}
L.A.R.J acknowledges the financial support from Brazilian Research Council FAP-DF grants $00193-00001247/2021-20$, $00193.00001808/2022-71$, and $00193-00000857/2021-14$, CNPq grants $302236/2018-0$ and 350176 /2022-1, and FAP-DF-PRONEM grant $00193.00001247/2021-20$. W.F.G acknowledges the financial support from FAP-DF grant $00193-00000811/2021-97$. This study was financed in part by the Coordenação de Aperfeiçoamento de Pessoal de Nível Superior – Brasil (CAPES) – Finance Code 88887.691997/2022-00. L.A.R.J. gratefully acknowledges the support from ABIN grant 08/2019.
 
\printcredits
\bibliographystyle{unsrt}
\bibliography{cas-refs}

\begin{thebibliography}{10}

\bibitem{huang2021two}
Hui Huang, Wei Feng, and Yu~Chen.
\newblock Two-dimensional biomaterials: material science, biological effect and
  biomedical engineering applications.
\newblock {\em Chemical Society Reviews}, 50(20):11381--11485, 2021.

\bibitem{tang2014two}
Hongjie Tang, Colin~M Hessel, Jiangyan Wang, Nailiang Yang, Ranbo Yu, Huijun
  Zhao, and Dan Wang.
\newblock Two-dimensional carbon leading to new photoconversion processes.
\newblock {\em Chemical Society Reviews}, 43(13):4281--4299, 2014.

\bibitem{chen2010graphene}
Da~Chen, Longhua Tang, and Jinghong Li.
\newblock Graphene-based materials in electrochemistry.
\newblock {\em Chemical Society Reviews}, 39(8):3157--3180, 2010.

\bibitem{khan2020recent}
Karim Khan, Ayesha~Khan Tareen, Muhammad Aslam, Renheng Wang, Yupeng Zhang,
  Asif Mahmood, Zhengbiao Ouyang, Han Zhang, and Zhongyi Guo.
\newblock Recent developments in emerging two-dimensional materials and their
  applications.
\newblock {\em Journal of Materials Chemistry C}, 8(2):387--440, 2020.

\bibitem{jia2017synthesis}
Zhiyu Jia, Yongjun Li, Zicheng Zuo, Huibiao Liu, Changshui Huang, and Yuliang
  Li.
\newblock Synthesis and properties of 2d carbon-graphdiyne.
\newblock {\em Accounts of chemical research}, 50(10):2470--2478, 2017.

\bibitem{bhimanapati2015recent}
Ganesh~R Bhimanapati, Zhong Lin, Vincent Meunier, Yeonwoong Jung, Judy Cha,
  Saptarshi Das, Di~Xiao, Youngwoo Son, Michael~S Strano, Valentino~R Cooper,
  et~al.
\newblock Recent advances in two-dimensional materials beyond graphene.
\newblock {\em ACS nano}, 9(12):11509--11539, 2015.

\bibitem{geim2007rise}
Andre~K Geim and Konstantin~S Novoselov.
\newblock The rise of graphene.
\newblock {\em Nature materials}, 6(3):183--191, 2007.

\bibitem{wu2012graphene}
Huang Wu and Lawrence~T Drzal.
\newblock Graphene nanoplatelet paper as a light-weight composite with
  excellent electrical and thermal conductivity and good gas barrier
  properties.
\newblock {\em Carbon}, 50(3):1135--1145, 2012.

\bibitem{papageorgiou2017mechanical}
Dimitrios~G Papageorgiou, Ian~A Kinloch, and Robert~J Young.
\newblock Mechanical properties of graphene and graphene-based nanocomposites.
\newblock {\em Progress in materials science}, 90:75--127, 2017.

\bibitem{nan2013thermal}
Hai~Yan Nan, Zhen~Hua Ni, Jun Wang, Zainab Zafar, Zhi~Xiang Shi, and Ying~Ying
  Wang.
\newblock The thermal stability of graphene in air investigated by raman
  spectroscopy.
\newblock {\em Journal of Raman Spectroscopy}, 44(7):1018--1021, 2013.

\bibitem{avouris2010graphene}
Phaedon Avouris.
\newblock Graphene: electronic and photonic properties and devices.
\newblock {\em Nano letters}, 10(11):4285--4294, 2010.

\bibitem{raccichini2015role}
Rinaldo Raccichini, Alberto Varzi, Stefano Passerini, and Bruno Scrosati.
\newblock The role of graphene for electrochemical energy storage.
\newblock {\em Nature materials}, 14(3):271--279, 2015.

\bibitem{shen2012biomedical}
He~Shen, Liming Zhang, Min Liu, and Zhijun Zhang.
\newblock Biomedical applications of graphene.
\newblock {\em Theranostics}, 2(3):283, 2012.

\bibitem{wan2011graphene}
Xiangjian Wan, Guankui Long, Lu~Huang, and Yongsheng Chen.
\newblock Graphene--a promising material for organic photovoltaic cells.
\newblock {\em Advanced Materials}, 23(45):5342--5358, 2011.

\bibitem{schwierz2010graphene}
Frank Schwierz.
\newblock Graphene transistors.
\newblock {\em Nature nanotechnology}, 5(7):487--496, 2010.

\bibitem{weiss2012graphene}
Nathan~O Weiss, Hailong Zhou, Lei Liao, Yuan Liu, Shan Jiang, Yu~Huang, and
  Xiangfeng Duan.
\newblock Graphene: an emerging electronic material.
\newblock {\em Advanced materials}, 24(43):5782--5825, 2012.

\bibitem{zhou2007substrate}
S~Yi Zhou, G-H Gweon, AV~Fedorov, de~First, PN, WA~De~Heer, D-H Lee, F~Guinea,
  AH~Castro~Neto, and A~Lanzara.
\newblock Substrate-induced bandgap opening in epitaxial graphene.
\newblock {\em Nature materials}, 6(10):770--775, 2007.

\bibitem{desyatkin2022scalable}
Victor~G Desyatkin, William~B Martin, Ali~E Aliev, Nathaniel~E Chapman,
  Alexandre~F Fonseca, Douglas~S Galv{\~a}o, Ericka~Roy Miller, Kevin~H Stone,
  Zhong Wang, Dante Zakhidov, et~al.
\newblock Scalable synthesis and characterization of multilayer
  $\gamma$-graphyne, new carbon crystals with a small direct band gap.
\newblock {\em Journal of the American Chemical Society}, 144(39):17999--18008,
  2022.

\bibitem{hou2022synthesis}
Lingxiang Hou, Xueping Cui, Bo~Guan, Shaozhi Wang, Ruian Li, Yunqi Liu, Daoben
  Zhu, and Jian Zheng.
\newblock Synthesis of a monolayer fullerene network.
\newblock {\em Nature}, 606(7914):507--510, 2022.

\bibitem{meirzadeh2023few}
Elena Meirzadeh, Austin~M Evans, Mehdi Rezaee, Milena Milich, Connor~J Dionne,
  Thomas~P Darlington, Si~Tong Bao, Amymarie~K Bartholomew, Taketo Handa,
  Daniel~J Rizzo, et~al.
\newblock A few-layer covalent network of fullerenes.
\newblock {\em Nature}, 613(7942):71--76, 2023.

\bibitem{fan2021biphenylene}
Qitang Fan, Linghao Yan, Matthias~W Tripp, Ond{\v{r}}ej Krej{\v{c}}{\'\i},
  Stavrina Dimosthenous, Stefan~R Kachel, Mengyi Chen, Adam~S Foster, Ulrich
  Koert, Peter Liljeroth, et~al.
\newblock Biphenylene network: A nonbenzenoid carbon allotrope.
\newblock {\em Science}, 372(6544):852--856, 2021.

\bibitem{pan2023long}
Fei Pan, Kun Ni, Tao Xu, Huaican Chen, Yusong Wang, Ke~Gong, Cai Liu, Xin Li,
  Miao-Ling Lin, Shengyuan Li, et~al.
\newblock Long-range ordered porous carbons produced from c60.
\newblock {\em Nature}, pages 1--7, 2023.

\bibitem{toh2020synthesis}
Chee-Tat Toh, Hongji Zhang, Junhao Lin, Alexander~S Mayorov, Yun-Peng Wang,
  Carlo~M Orofeo, Darim~Badur Ferry, Henrik Andersen, Nurbek Kakenov, Zenglong
  Guo, et~al.
\newblock Synthesis and properties of free-standing monolayer amorphous carbon.
\newblock {\em Nature}, 577(7789):199--203, 2020.

\bibitem{hu2022synthesis}
Yiming Hu, Chenyu Wu, Qingyan Pan, Yinghua Jin, Rui Lyu, Vikina Martinez,
  Shaofeng Huang, Jingyi Wu, Lacey~J Wayment, Noel~A Clark, et~al.
\newblock Synthesis of $\gamma$-graphyne using dynamic covalent chemistry.
\newblock {\em Nature Synthesis}, 1(6):449--454, 2022.

\bibitem{liu2022electronic}
Guogang Liu, Tong Chen, Xiaohui Li, Zhonghui Xu, and Xianbo Xiao.
\newblock Electronic transport in biphenylene network monolayer: Proposals for
  2d multifunctional carbon-based nanodevices.
\newblock {\em Applied Surface Science}, 599:153993, 2022.

\bibitem{chen2023new}
Hsin-Tsung Chen and Dinesh Kumar~Dhanthala Chittibabu.
\newblock A new carbon allotrope: Biphenylene as promising anode materials for
  li-ion and lio2 batteries.
\newblock {\em Solid State Ionics}, 395:116214, 2023.

\bibitem{ferguson2017biphenylene}
David Ferguson, Debra~J Searles, and Marlies Hankel.
\newblock Biphenylene and phagraphene as lithium ion battery anode materials.
\newblock {\em ACS applied materials \& interfaces}, 9(24):20577--20584, 2017.

\bibitem{mu2023superconducting}
Yuewen Mu and Si-Dian Li.
\newblock Superconducting and transport properties of biphenylene network
  monolayer with alkaline metal adatoms.
\newblock {\em Applied Surface Science}, page 157255, 2023.

\bibitem{esfandiarpour2022carbon}
Razieh Esfandiarpour, Fatemeh Zamanian, Farideh Badalkhani-Khamseh, and
  Mohammad~Reza Hosseini.
\newblock Carbon dioxide sensor device based on biphenylene nanotube: A density
  functional theory study.
\newblock {\em Computational and Theoretical Chemistry}, 1218:113939, 2022.

\bibitem{su2022theoretical}
Wan-Sheng Su and Chen-Hao Yeh.
\newblock Theoretical investigation of methane oxidation reaction over a novel
  metal-free catalyst biphenylene network.
\newblock {\em Diamond and Related Materials}, 124:108897, 2022.

\bibitem{mishra2011carbon}
Ashish~Kumar Mishra and Sundara Ramaprabhu.
\newblock Carbon dioxide adsorption in graphene sheets.
\newblock {\em AIP Advances}, 1(3):032152, 2011.

\bibitem{cabrera2009adsorption}
Pepa Cabrera-Sanfelix.
\newblock Adsorption and reactivity of co2 on defective graphene sheets.
\newblock {\em The Journal of Physical Chemistry A}, 113(2):493--498, 2009.

\bibitem{takeuchi2017adsorption}
Kaori Takeuchi, Susumu Yamamoto, Yuji Hamamoto, Yuichiro Shiozawa, Keiichiro
  Tashima, Hirokazu Fukidome, Takanori Koitaya, Kozo Mukai, Shinya Yoshimoto,
  Maki Suemitsu, et~al.
\newblock Adsorption of co2 on graphene: a combined tpd, xps, and vdw-df study.
\newblock {\em The Journal of Physical Chemistry C}, 121(5):2807--2814, 2017.

\bibitem{delley_JCP}
B.~Delley.
\newblock An all‐electron numerical method for solving the local density
  functional for polyatomic molecules.
\newblock {\em The Journal of Chemical Physics}, 92(1):508--517, 1990.

\bibitem{delley_JCP2}
B.~Delley.
\newblock From molecules to solids with the dmol3 approach.
\newblock {\em The Journal of Chemical Physics}, 113(18):7756--7764, 2000.

\bibitem{systemes2017biovia}
Dassault Syst{\`e}mes.
\newblock Biovia materials studio.
\newblock {\em San Diego}, 2017.

\bibitem{grimme2006}
Stefan Grimme.
\newblock Semiempirical gga-type density functional constructed with a
  long-range dispersion correction.
\newblock {\em Journal of computational chemistry}, 27(15):1787--1799, 2006.

\bibitem{kresse1999}
Georg Kresse and Daniel Joubert.
\newblock From ultrasoft pseudopotentials to the projector augmented-wave
  method.
\newblock {\em Physical review b}, 59(3):1758, 1999.

\bibitem{perdew1996}
John~P Perdew, Kieron Burke, and Matthias Ernzerhof.
\newblock Generalized gradient approximation made simple.
\newblock {\em Physical review letters}, 77(18):3865, 1996.

\bibitem{delley2002}
B~Delley.
\newblock Hardness conserving semilocal pseudopotentials.
\newblock {\em Physical Review B}, 66(15):155125, 2002.

\bibitem{hendrik1976}
Hendrik~J Monkhorst and James~D Pack.
\newblock Special points for brillouin-zone integrations.
\newblock {\em Physical review B}, 13(12):5188, 1976.

\bibitem{lima2021}
Kleuton A~Lopes Lima, Marcelo L~Pereira J{\'u}nior, F{\'a}bio~F Monteiro,
  Luiz~F Roncaratti, and Luiz A~Ribeiro J{\'u}nior.
\newblock O2 adsorption on defective penta-graphene lattices: A dft study.
\newblock {\em Chemical Physics Letters}, 763:138229, 2021.

\bibitem{tlima2019}
Igo~T Lima, Ricardo Gargano, Silvete Guerini, and Edson~NC Paura.
\newblock A theoretical study of adsorbed non-metallic atoms on magnesium
  chloride monolayers.
\newblock {\em New Journal of Chemistry}, 43(20):7778--7783, 2019.

\bibitem{paura2014}
Edson~NC Paura, Wiliam~F da~Cunha, Jo{\~a}o Batista~Lopes Martins,
  Geraldo~Magela e~Silva, Luiz~F Roncaratti, and Ricardo Gargano.
\newblock Carbon dioxide adsorption on doped boron nitride nanotubes.
\newblock {\em Rsc Advances}, 4(54):28249--28258, 2014.

\bibitem{mananghaya2020adsorption}
Michael~Rivera Mananghaya.
\newblock Adsorption of co and desorption of co2 interacting with pt (111)
  surface: a combined density functional theory and kinetic monte carlo
  simulation.
\newblock {\em Adsorption}, 26(3):461--469, 2020.

\bibitem{lima2019}
Kleuton Antunes~Lopes Lima, Wiliam Ferreira~da Cunha, F{\'a}bio~Ferreira
  Monteiro, Bernhard~Georg Enders, Marcelo Lopes~Pereira Jr, and Luiz
  Antonio~Ribeiro Jr.
\newblock Adsorption of carbon dioxide and ammonia in transition metal--doped
  boron nitride nanotubes.
\newblock {\em Journal of Molecular Modeling}, 25:1--7, 2019.

\bibitem{bafekry2021biphenylene}
A~Bafekry, M~Faraji, MM~Fadlallah, HR~Jappor, S~Karbasizadeh, M~Ghergherehchi,
  and D~Gogova.
\newblock Biphenylene monolayer as a two-dimensional nonbenzenoid carbon
  allotrope: a first-principles study.
\newblock {\em Journal of Physics: Condensed Matter}, 34(1):015001, 2021.

\bibitem{mortazavi2022anisotropic}
Bohayra Mortazavi and Alexander~V Shapeev.
\newblock Anisotropic mechanical response, high negative thermal expansion, and
  outstanding dynamical stability of biphenylene monolayer revealed by
  machine-learning interatomic potentials.
\newblock {\em FlatChem}, 32:100347, 2022.

\bibitem{rahaman2017metamorphosis}
Obaidur Rahaman, Bohayra Mortazavi, Arezoo Dianat, Gianaurelio Cuniberti, and
  Timon Rabczuk.
\newblock Metamorphosis in carbon network: From penta-graphene to biphenylene
  under uniaxial tension.
\newblock {\em FlatChem}, 1:65--73, 2017.

\bibitem{sanyal2009molecular}
Biplab Sanyal, Olle Eriksson, Ulf Jansson, and Helena Grennberg.
\newblock Molecular adsorption in graphene with divacancy defects.
\newblock {\em Physical Review B}, 79(11):113409, 2009.

\bibitem{timsorn_IOP}
Kriengkri Timsorn and Chatchawal Wongchoosuk.
\newblock Adsorption of {NO}2, {HCN}, {HCHO} and {CO} on pristine and amine
  functionalized boron nitride nanotubes by self-consistent charge density
  functional tight-binding method.
\newblock {\em Materials Research Express}, 7(5):055005, may 2020.

\end{thebibliography}

\end{document}